\documentclass[11pt]{article}
\usepackage{amssymb}
\usepackage{graphics}
\usepackage{epsfig}
\begin{document}
\title{ Higgs boson pair production in the little Higgs model
at hadron colliders \footnote{Supported by National Natural
Science Foundation of China.}} \vspace{3mm}

\author{{ Liu Jing-Jing$^{2}$, Ma Wen-Gan$^{1,2}$, Li Gang$^{2}$, Zhang Ren-You$^{2}$ and Hou Hong-Sheng$^{2}$}\\
{\small $^{1}$ CCAST (World Laboratory), P.O.Box 8730, Beijing 100080, P.R.China} \\
{\small $^{2}$ Department of Modern Physics, University of Science and Technology}\\
{\small of China (USTC), Hefei, Anhui 230027, P.R.China} }

\date{}
\maketitle \vskip 12mm

\begin{abstract}
The Higgs boson pair production process at hadron collider
provides an opportunity for performing a study of the trilinear
Higgs boson self-coupling. In this paper, we analyze the pair
production of a neutral Higgs boson via both gluon-gluon and
$b$-$\bar{b}$ fusions in the littlest Higgs (LH) model at the CERN
LHC. We find that in some parameter space the relative corrections
of the total cross section to the SM prediction may reach a value
of $24\%$ when $x~(=4 f v'/v^2)=0.95$ at the LHC. We conclude that
if the parameter $x$ has a value above 0.8, the relative
corrections contributed by the LH model reach values beyond $8\%$
and can be observed at the LHC.

\end{abstract}


\vfill \eject

\baselineskip=0.36in

\newcommand{\nb}{\nonumber}
\newcommand{\mch}{m_{\tilde{\chi}^+_i}}
\newcommand{\mno}{m_{\tilde{\chi}^0_j}}

\section{Introduction}

\par
The standard model(SM) \cite{s1,s2} theory has been proved by all
existing precise experimental data with its theoretical
predictions beyond one-loop level being coincident with
experimental observations. But in the SM the Higgs boson mass
suffers from an instability under radiative corrections. This
"hierarchy problem" motivates much of current research works about
new physics beyond the SM. Among the extended models beyond the
SM, the little Higgs model offers a very promising solution to the
hierarchy problem in which the Higgs boson is naturally light as a
result of nonlinearly realized symmetry \cite{LH1}-\cite{LH6}. The
first successful model, which cancels all relevant quadratic
divergences based on the pseudo Goldstone idea, was constructed by
Arkani-Hamed, Cohen, and Georgi \cite{LH1}. Then more models were
constructed, such as $SU(5)/SO(5)$ \cite{LH5}, $SU(6)/SP(6)$
\cite{LH4}, and the minimal moose $SU(3)^2/SU(3)$ \cite{LH3} and
general moose $SU(3)^n/SU(3)^k$ \cite{LH7}. The most economical
model of them is the littlest Higgs (LH) model, which is based on
an $SU(5)/SO(5)$ nonlinear sigma model \cite{LH5}. It consists of
a $SU(5)$ global symmetry, which is spontaneously broken down to
$SO(5)$ by a vacuum condensate $f$. In the LH model, a set of new
heavy gauge bosons ($A_H,Z_H,W_H$) and a new heavy-vector-like
quark ($T$) are introduced which just cancel the quadratic
divergence induced by SM gauge boson loops and the top quark loop,
respectively.

\par
One of the most important task of  present and future experiments
is to search for Higgs boson and investigate its properties.
Studying the properties of the Higgs potential will reveal details
of mass-generation mechanism in spontaneously broken gauge
theories, which can be obtained through measuring the Higgs boson
self-interactions. Multiple Higgs boson production processes at
hadron colliders provide the way to probe the Higgs boson
self-interactions. Many works have been contributed to studies of
Higgs-pair production at the hadron collider in some traditional
models \cite{L1}-\cite{L4}. The possibility of measuring the Higgs
boson self-coupling at the LHC has been investigated by U. Baur,
T. Plehn and D. Rainwater \cite{Baur}. They found that it should
be possible at the CERN Large Hadron Collider (LHC) with design
luminosity to establish that the SM Higgs boson has nonzero self-
coupling and that $\lambda/\lambda_{SM}$ can be restricted to a
range of $0-3.7$ at $95\%$ confidence level if its mass is between
$150$ to $200$ GeV. Recently, the LH model contribution to the
Higgs decay width was investigated in Refs. \cite{Dib} and
\cite{LH9}. Dib \textit{et al.}, discussed also the LH model
contribution to the process $pp \to H^0H^0+X$ in Ref. \cite{Dib}.
There they did not consider mixing and interference effects
between the SM particles and the new heavy states, and thus they
got negligible results of the order of $(v/f)^4$ and concluded
that the contribution from the LH model to the pair production to
the Higgs bosons seems to be unobservable at the LHC \cite{Dib}.
If the interference and mixing effects are included in the
analysis, the contribution is at the order $(v/f)^2$ and does
change obviously the results as compared to the SM prediction in
some parameter space \cite{LH9}.

\par
In this paper we investigate the effect of the LH model
 on neutral Higgs boson pair production via both gluon and bottom fusions at the CERN Large
Hadron Collider (i.e., $g g \to H^0 H^0$ and $b \bar{b}\to H^0
H^0$ ) at the complete lowest order, considering mixing and
interference effects between the SM particles and the new heavy
states. The paper is organized as follows. In Sec. 2, we briefly
go through the LH model theory. In Sec. 3, we present an
analytical evaluation. The numerical results, discussion and a
short summary are given in Sec. 4. Finally we present the relevant
Feynman rules in the Appendix.

\vskip 10mm
\section{littlest Higgs model}
\par
The littlest Higgs model is based on an $SU(5)/SO(5)$ nonlinear
sigma model. At the scale $\Lambda_s \sim 4 \pi f$, the vacuum
expectation value (VEV) associated with the spontaneous symmetry
breaking proportional to the scale $f$ is parameterized by the $5
\times 5$ symmetry matrix \cite{LH5} \cite{LH8}
\begin{equation}
\Sigma_0 = \left( \begin{array}{ccc}
 & & {\bf{1}}_{2 \times 2} \\
 &1 & \\
{\bf{1}}_{2 \times 2} & &
\end{array}\right).
\end{equation}
The VEV breaks the $SU(5)$ global symmetry into its subgroup
$SO(5)$ and breaks the local gauge symmetry $[SU(2) \otimes
U(1)]^2$ into its diagonal subgroup $SU(2)_L \otimes U(1)_Y$ at
the same time, which is identified as the SM electroweak gauge
group. The scalar fields are parameterized by
\begin{eqnarray}
\Sigma(x) = e^{i \Pi(x)/f} \Sigma_0 e^{i \Pi(x)^{T}/f},
\end{eqnarray}
where $\Pi(x) = \pi^a(x) X^a$ is the Goldstone boson matrix. $X^a$
are the broken generators of $SU(5)$ which obey the relation
\begin{equation}
X^a\Sigma_0-\Sigma_0 X^{aT}=0.
\end{equation}
Therefore, the Goldstone boson matrix $\Pi(x)$ can be expressed as
\begin{equation}
\Pi=\left( \begin{array}{ccc} {}& h^\dagger/\sqrt 2&
\phi^\dagger\\h/\sqrt 2 &{}& h^*/\sqrt 2\\\phi  & h^T/\sqrt 2 &{}
    \end{array}\right),
\end{equation}
where
\begin{eqnarray}
h = \left(
    \begin{array}{cc}
    h^{+} ~~ h^0
    \end{array}
    \right), ~~~~~~
\phi = \left(
       \begin{array}{cc}
       \phi^{++} & \phi^{+}/\sqrt{2} \\
       \phi^{+}/\sqrt{2} & \phi^0
       \end{array}
       \right)
\end{eqnarray}
are a doublet and triplet under the unbroken $SU(2)_L \otimes
U(1)_Y$ SM gauge group, respectively.

\par
The leading-order dimension-2 term for the scalar fields
$\Sigma(x)$ in the littlest Higgs model can be written as
\begin{equation}
{\cal L} = \frac{1}{2} \frac{f^2}{4} {\rm Tr} |{\cal D}_{\mu}
\Sigma |^2.
\end{equation}
${\cal D}_{\mu}$ is the covariant derivative for gauge group
$[SU(2) \otimes U(1)]^2 = [SU(2)_1 \otimes U(1)_1] \otimes
[SU(2)_2 \otimes U(1)_2]$. It is defined as
\begin{equation}
{\cal D}_\mu \Sigma =
\partial_\mu \Sigma - i \sum_{j=1}^2
\left[ g_j ( W_j \Sigma + \Sigma W_j^T) + g_j^{\prime} (B_j \Sigma
+ \Sigma B_j^T) \right],
\end{equation}
where $W_{\mu j} = \sum_{a = 1}^3 W_{\mu j}^a Q_j^a$ and $B_j =
B_{\mu j} Y_j$ are the $SU(2)_j$ and $U(1)_j$ gauge fields,
respectively. The generators of two $SU(2)$'s ($Q_j^a$) and two
$U(1)$'s generators ($Y_j$) are as follows
\begin{eqnarray}
Q_1^a = \left(
        \begin{array}{cc}
        \frac{\sigma^a}{2} & \\
        & {\bf{0}}_{3 \times 3}
        \end{array}
    \right),&~~&
Q_2^a = \left(
        \begin{array}{cc}
        {\bf{0}}_{3 \times 3} & \\
        & -\frac{\sigma^{a \ast}}{2}
\end{array}\right), \nonumber \\
Y_1 = {\rm diag}\{-3,~ -3,~ 2,~ 2,~ 2\}/10, &~~& Y_2 = {\rm
diag}\{-2,~ -2,~ -2,~ 3,~ 3\}/10,
\end{eqnarray}
where $\sigma^a~ (a = 1, 2, 3)$ are the Pauli matrices. As we
expect, the breaking of the gauge symmetry $[SU(2) \times U(1)]^2$
into its diagonal subgroup $SU(2)_L \times U(1)_Y$ gives rise to
heavy gauge bosons $W^{\prime}$ and $B^{\prime}$, and the
remaining unbroken subgroup $SU(2)_L \times U(1)_Y$ introduces the
massless gauge bosons $W$ and $B$.

\par
As we know, in the littlest Higgs model there is no Higgs
potential at the tree level. Instead, the Higgs potential is
generated at one loop and higher orders due to the interactions
with gauge bosons and fermions. Up to operators of dimension 4,
the Higgs potential (Coleman-Weinberg potential) can be expressed
as \cite{LH9} \cite{LH8}
\begin{eqnarray}
\label{win} V &=& \lambda_{\phi^2} f^2 {\rm Tr}( \phi^{\dag} \phi)
+ i \lambda_{h \phi h} f (h \phi^{\dag} h^T - h^{\ast} \phi
h^{\dag}) - \mu^2 h h^{\dag} +
\lambda_{h^4} (h h^{\dag})^2 \nonumber \\
& & + \lambda_{h \phi \phi h} h \phi^{\dag} \phi h^{\dag} +
\lambda_{h^2 \phi^2} h h^{\dag} {\rm Tr}(\phi^{\dag} \phi) +
\lambda_{\phi^2 \phi^2} \left[
                        {\rm Tr}(\phi^{\dag} \phi)
            \right]^2 +
\lambda_{\phi^4} {\rm Tr}(\phi^{\dag} \phi \phi^{\dag} \phi),
\end{eqnarray}
where $\lambda_{\phi^2}$, $\lambda_{h \phi h}$, $\lambda_{h^4}$,
$\lambda_{h \phi \phi h}$, $\lambda_{h^2 \phi^2}$,
$\lambda_{\phi^2 \phi^2}$, and $\lambda_{\phi^4}$ are the
coefficients of the original Higgs potential. The coefficients
which are concerned in our calculation have the expressions as
\cite{LH9}
\begin{eqnarray}
\label{win1}\lambda_{\phi^2}= \frac{{M_\phi}^2}{f^2},~
\lambda_{h\phi h}=\frac{x {M_\phi}^2}{2f^2},
~\lambda_{h^4}=\frac{{M_\phi}^2}{4 f^2}, ~\lambda_{h\phi \phi
h}=-\frac{4{M_\phi}^2}{3f^2}.
\end{eqnarray}
By minimizing the Coleman-Weinberg potential, we obtain the vacuum
expectation values $\langle h^0 \rangle =v/\sqrt 2$, $\langle i
\phi^0 \rangle=v^{\prime}$ of the Higgs boson doublet $h$ and
triplet $\phi$, which give rise to electroweak symmetry breaking
(EWSB). The Coleman-Weinberg potential provides the trilinear and
quartic Higgs self-couplings. After EWSB, the gauge sector gets an
additional mass and mixing term due to the VEVs of $h$ and $\phi$.
By diagonalizing the quadratic term of the gauge sector, we may
get the mass eigenstates $A_L$, $Z_L$, $W_L$, $A_H$, $Z_H$, and
$W_H$ and their masses.

\par
To avoid large quadratic divergence in the Higgs boson mass due to
the heavy top quark Yukawa interaction, we introduce a pair of new
fermions $\tilde{t}$ and $\tilde{t}^{\prime}$ \cite{LH5}
\cite{LH8} and a set of new interactions. The scalar couplings to
the top quark can be taken from the following Lagrangian
\cite{LH5} \cite{LH8}:
\begin{equation}
\label{L1}{\cal L}_Y = \frac{1}{2}\lambda_1 f
\epsilon_{ijk}\epsilon_{xy}\chi_i \Sigma_{jx}\Sigma_{ky}u^{\prime
c}_3+ \lambda_2f \tilde{t}\tilde{t}^{\prime c}+H.C.
\end{equation}
where $\chi=(b_3,~ t_3,~\tilde{t})$, $\epsilon_{ijk}$ and
$\epsilon_{xy}$ are antisymmetric tensors where $i$, $j$, $k$ run
through 1, 2, 3 and $x$, $y$ run through 4, 5, and $\lambda_1$ and
$\lambda_2$ are the new model parameters. By expanding the above
Lagrangian, we get the physical states of the top quark $t$ and a
new heavy-vector-like quark $T$. The masses of the two physical
states are
\begin{eqnarray}
\label{mt1} m_t = \frac{ v \lambda_1\lambda_2 }{\sqrt{\lambda_1^2
  + \lambda_2^2}} \left\{ 1+\frac{v^2}{f^2} \left [-\frac{1}{3}+\frac{fv'}{v^2}+
  \frac{1}{2}\frac{\lambda_1^2}{\lambda_1^2+\lambda_2^2} \left(
  1-\frac{\lambda_1^2}{\lambda_1^2+\lambda_2^2} \right ) \right ] \right\} ~ ,
\end{eqnarray}
\begin{eqnarray}
\label{mt2}m_T = f\sqrt{\lambda_1^2 + \lambda_2^2}\left[ 1+{\cal
O}(v^2/f^2) \right],
\end{eqnarray}
respectively. Since the top quark mass is already obtained in the
SM, we can then get the parameter relation from Eq. (\ref{mt1}) as
deduced in Ref. \cite{LH8}
\begin{eqnarray}
\label{l1l2}\frac{1}{\lambda_1^2}+\frac{1}{\lambda_2^2}\approx
\frac{v^2}{{m_t}^2} \approx 2.
\end{eqnarray}

\vskip 10mm
\section{Calculation}
\par
At hadron collider, the Higgs boson pair can be produced through
two mechanisms. One is loop-induced production via gluon fusion;
the other is from $b\bar{b}$ annihilation. The Feynman diagrams
contributing to the subprocess $gg \to H^0H^0$, which are involved
in the framework of the LH model, are depicted in Fig. 1. The
diagrams created by exchanging two external gluon lines and final
Higgs boson lines are not shown there. In Fig. 1, the first three
diagrams(Fig. 1(1)-1(3)) [excluding the Fig. 1(1) with a heavy-
vector-like quark ($T$) loop] are just the same as those in the
framework of the SM, while the remaining four figures (Fig.
1(4)-1(7)) are the extra diagrams beyond the SM. All the Feynman
diagrams can be classified into three types. The first type is
named as \textit{s}-channel diagrams with the exchange of a
virtual neutral Higgs boson $H^0$ or heavy triplet Higgs boson
$\Phi^0$ which couples to a pair of gluons via a triangle quark
loop [shown in Fig.1(1), 1(4)]. The second type is called box
diagrams [shown in Fig. 1(2), 1(3), 1(5), 1(6)] and the third is
the quartic interaction type where the neutral Higgs bosons are
produced by means of quartic interactions [shown in Fig. 1(7)].
All the relevant Feynman rules can be found in Ref. \cite{LH8} and
the Appendix in this paper. In the loop diagram calculation of
this subprocess, we adopt a dimensional regularization scheme. The
Feynman diagrams for the subprocess $b\bar{b} \to H^0 H^0 $ are
depicted in Fig. 2. In this work we adopted the Feynman-'t Hooft
gauge.

\par
In our calculation, we denote the two subprocesses as
\begin{eqnarray}
  g(c_1,~p_1) + g(c_2,~p_2) &\to& H^0(k_1) + H^0(k_2)~,\\
  b(c_3,~p_1) + \bar{b}(c_4,~p_2) &\to& H^0(k_1) + H^0(k_2)~.
\end{eqnarray}
where $c_1$, $c_2$, $c_3$, and $c_4$ are the color indexes of the
two initial particles ($gg$ and $b \bar{b}$). $p_1$, $p_2$ and
$k_1$, $k_2$ are the incoming and outgoing four-momenta of the
initial and final particles, respectively. In the subprocess $gg
\to H^0H^0$, we define $\theta$ as the scattering angle between
one of the gluons and one of the final $H^0$ bosons, and in the
subprocess $b\bar{b}\to H^0H^0$, $\theta$ represents the
scattering angle between the $b$ quark and one of the final Higgs
bosons. In the center-of-mass system (c.m.s.) the four-momenta of
the initial and final particles can be expressed as $\footnote{In
our calculation we set the mass of the $b$ quark to be zero except
in the Yukawa coupling.}$

\begin{eqnarray}
p_{1}&=&(\frac{\sqrt{\hat{s}}}{2},~0,~0,~\frac{\sqrt{\hat{s}}}{2}), \nonumber \\
p_{2}&=&(\frac{\sqrt{\hat{s}}}{2},~0,~0,~-\frac{\sqrt{\hat{s}}}{2}), \nonumber \\
k_{1}&=&(\frac{\sqrt{\hat{s}}}{2},~ \frac{\sqrt{\hat{s}}}{2} \beta
_{H^{0}}\sin \theta, ~~ 0, ~\frac{\sqrt{\hat{s}}}{2}\beta
_{H^{0}}\cos \theta),\nonumber\\
k_{2}&=&(\frac{\sqrt{\hat{s}}}{2},~ -\frac{\sqrt{\hat{s}}}{2}
\beta _{H^{0}}\sin \theta, ~~ 0, ~-\frac{\sqrt{\hat{s}}}{2}\beta
_{H^{0}}\cos \theta),
\end{eqnarray}
where $\beta_{H^{0}}=\sqrt{1-4m^{2}_{H^{0}}/\hat{s}}$ is the
velocity of final neutral Higgs bosons in the c.m.s. and $\hat{s}
= (p_1 + p_2)^2$.

\par
The amplitude for the subprocess $g g \rightarrow H^0 H^0$ can be
expressed as
\begin{eqnarray}
{\cal M}\left(g g \to H^0 H^0\right)
&=&{\cal M}^{(t)}+{\cal M}^{(b)}+{\cal M}^{(q)} \nonumber \\
&=& \epsilon_{\mu}(p_{1})\epsilon_{\nu}(p_{2}) \left(f_1
g^{\mu\nu}\delta_{c_1c_2} +f_2{k_1}^\mu {k_1}^\nu\delta_{c_1c_2}
+f_3\epsilon^{\alpha\mu\nu\beta}k_{1\alpha} p_{1\beta} \right.\nonumber\\
& & \left.+f_4\epsilon^{\alpha\mu\nu\beta}k_{1\alpha}p_{2\beta}
 +f_5\epsilon^{\mu\nu\alpha\beta}p_{1\alpha}p_{2\beta}
\right),
\end{eqnarray}
where ${\cal M}^{(t)}$, ${\cal M}^{(b)}$, and ${\cal M}^{(q)}$
represent the amplitudes of triangle, box, and quartic diagrams,
respectively.

\par
For the subprocess $b \bar{b} \rightarrow H^0 H^0$, the amplitude
can be expressed as
\begin{eqnarray}
{\cal M}(b\bar{b} \rightarrow H^0H^0) &=& \bar{v}\left(p_{2})
(g_{1}+g_{2} \rlap/k_{1}\right)\delta_{c_3c_4}u(p_{1}).
\end{eqnarray}
In the above two equations $f_i~ (i = 1,...,5)$ and $g_j~ (j = 1,
2)$ are the form factors of the two subprocesses, respectively.
Since the explicit expressions of these form factors are lengthy,
we do not list them in this paper.\footnote{The Mathematica
program codes of all the form factors for $gg \to H^0H^0$ and
$b\bar{b} \to H^0H^0$ are obtainable by sending an email to
moonbank@mail.ustc.edu.cn}

\par
Then the total cross sections for these two subprocesses can be
written as
\begin{eqnarray}
\label{lab1} \hat{\sigma}(\hat{s},~ g g \rightarrow H^{0}H^{0})
&=& \frac{1}{32\pi\hat{s}^{2}} \int^{\hat{t}^{+}}_{\hat{t}^{-}}d
\hat{t} \overline{\sum}
\mid {\cal M}(gg\rightarrow H^{0}H^{0}) \mid^{2}, \\
\label{lab2} \hat{\sigma}(\hat{s},~ b \bar{b}\rightarrow
H^{0}H^{0}) &=& \frac{1}{32\pi\hat{s}^{2}}
\int^{\hat{t}^{+}}_{\hat{t}^{-}}d \hat{t} \overline{\sum} \mid
{\cal M}(b\bar{b}\rightarrow H^{0}H^{0})\mid^{2},
\end{eqnarray}
respectively, where the bar over the summation recalls averaging
over initial spins and colors,
$\hat{t}^{\pm}=(m^{2}_{H^{0}}-\hat{s}/2 \pm
\hat{s}\beta_{H^{0}}/2)$, and due to the identical final two Higgs
bosons, the right-hand sides of Eqs. (\ref{lab1}) and (\ref{lab2})
have been multiplied by an additional factor of $1/2$ separately.
The color and spin average factors for the subprocess $gg\to
H^0H^0$ are $1/64$ and $1/4$ and for the subprocess $b\bar{b}\to
H^0H^0$ are $1/9$ and $1/4$, respectively. The total cross section
for the neutral Higgs boson pair production through gluon (or
bottom) fusion in proton-proton collisions can be obtained by
performing the following integration
\begin{eqnarray}
\sigma(pp\to g g(b\bar{b})\to H^0H^0) =
\int^{1}_{4{m^2_{H^{0}}}/s} d\tau\frac{dL_{ij}}{d\tau}\hat{\sigma}
(\hat{s} = \tau s,~ gg(b\bar{b})\to H^{0}H^{0}),
\end{eqnarray}
where $\sqrt{s}$ and $\sqrt{\hat{s}}$ denote the $pp$ and $g g$
(or $b \bar{b}$) c.m.s. energies, respectively, and $dL_{ij}/d\tau
$ is the luminosity of colliding partons, which is defined as
\begin{eqnarray}
\frac{dL_{ij}}{d\tau}~=\frac{1}{1+\delta_{ij}}~\int^1_\tau\frac{dx_1}{x_1}
\left[F_{i/p}(x_1,\mu)F_{j/p}(\frac{\tau}x_1{,\mu}) +(i
\leftrightarrow j)\right].
\end{eqnarray}
In our calculation we adopt the CTEQ6 parton distribution function
\cite{cacu1} and take the factorization scale $\mu$ to be $2
m_{H^0}$ in the subprocess $gg \to H^0H^0$, while $\mu$ to be
$m_{H^0}/2$ in the calculation of the subprocess $b\bar{b}\to
H^0H^0$ \cite{L5}. The numerical calculation is carried out for
the LHC at an energy of $14~{\rm TeV}$.

\vskip 10mm
\section{Numerical results and discussions }
\par
In the numerical evolution we take the input parameters as $m_W =
80.423~{\rm GeV}$, $m_Z = 91.1876~{\rm GeV}$, $m_u=4.5~{\rm MeV}$,
$m_d=8.5~{\rm MeV}$, $ m_s=150~{\rm MeV}$, $m_c=1.25~{\rm GeV}$,
$m_t=174.3~{\rm GeV}$, and $\alpha(m_Z)=1/128$. We use a simple
one-loop formula to express the running strong coupling constant
$\alpha_s$:
\begin{eqnarray}
\alpha_{s}(\mu)=\frac{\alpha_{s}(m_{Z})}{1+\frac{33-2n_{f}}
{6\pi}\alpha_{s}(m_{Z})ln(\frac{\mu}{m_{Z}})},
\end{eqnarray}
where $\alpha_{s}(m_{Z})=0.118$ and $n_f$ is the number of active
flavors at scale $\mu$ \cite{data}.
\par
In the numerical calculation, we use the next-to-leading order
formula to evaluate the running mass of bottom quark
$\overline{m}_b(Q)$ \cite{mass}.
\begin{eqnarray}
\overline{m}_b(Q)&=&U_5(Q,\overline{m}_b)\overline{m}_b(\overline{m}_b),
~~~~~~~~~~~~~~~~~~~~~~~~~Q<m_t,\nonumber\\
\overline{m}_b(Q)&=&U_6(Q,m_t)U_5(m_t,\overline{m}_b)\overline{m}_b(\overline{m}_b),
~~~~~~~~~~~Q>m_t,
\end{eqnarray}
where $\overline{m}_b=\overline{m}_b(\overline{m}_b)=4.25$ GeV and
the energy scale $Q$ is taken to be $2 m_{H^0}$ in our
calculation. The evolution factor $U_f(f=5,6)$ is
\begin{eqnarray}
U_f(Q_2,Q_1)&=&\left(\frac{\alpha_s(Q_2)}{\alpha_s(Q_1)}\right)^{d^f}
\left[1+\frac{\alpha_s(Q_1)-\alpha_s(Q_2)}{4\pi}J^f\right],\nonumber\\
d^f&=&\frac{12}{32-2 f}, \nonumber\\
J^f&=&-\frac{8982-504 f+40 f^2}{3(33-2 f)^2}.
\end{eqnarray}
\par
In the LH model, the calculation of the Higgs boson pair
production at hadron colliders involves four additional free
parameters. One is the parameter $f$, which is the symmetry
breaking scale parameter at TeV order. The direct and indirect
effects of the LH model provided by the present experimental
measurements have placed a constraint $f \gtrsim 3.5~{\rm TeV}$,
although it depends on the model assumption about the $U(1)'s$
\cite{LH8} \cite{LH10} \cite{LH11}. The second one is the mass of
the Higgs boson, $m_{H^0}$. The third parameter is the coefficient
$\lambda_2$, which is the coupling constant of the new heavy-
vector-like quark $T$. Because there is a relation between the top
quark and heavy-vector-like quark coupling parameters $\lambda_1$
and $\lambda_2$, as shown in Eq. (\ref{l1l2}), we can use
$\lambda_1/\lambda_2$ to parametrize the mass of the new heavy
vector quark [see Eq. (\ref{mt2})]. The last one is $v'$, and we
define $x=4 f v'/v^2$ to parametrize this vacuum expectation value
of the scalar triplet field $\phi$. The masses of neutral scalar
boson $M_{\Phi^0}$ can be given as \cite{LH9} \cite{LH8}
\begin{eqnarray}
M_{\Phi^0}^2~=~\frac{2m_{H^0}^2f^2}{v^2[1-(4v'f/v^2)^2]}=~\frac{2m_{H^0}^2f^2}{v^2(1-x^2)}~.
\end{eqnarray}
The above equation about the mass of $\Phi$ requires a constraint
of $0 \leq x<1$ (i.e., $4v'f/v^2 \lesssim 1$), which shows the
relation between the scale $f$ and the vacuum expectation values
of the Higgs field doublet and triplet ($v$,$v'$). Based on the
limitation of current electroweak experimental data \cite{LH8}
\cite{LH10} \cite{LH11}, in our calculation we choose $f=3.5~{\rm
TeV}$ unless otherwise stated.

\par
Since our numerical calculation shows that the contributions of
subprocess $b\bar{b} \to H^0H^0$ to the parent process $pp\to
H^0H^0+X$ at the LHC is less than $1\%$ of the contributions from
subprocess $gg \to H^0H^0$, we present only plots of the
subprocess $gg \to H^0H^0$ for representation. In fact, our
analysis also demonstrates that the contributions of the
additional diagrams involving the new heavy-vector-like quark $T$
or neutral heavy triplet Higgs boson $\Phi^0$ for the subprocess
$gg \to H^0H^0$ are very small. The deviations of the cross
section from the SM are mainly aroused by the contributions of the
diagrams which exist also in the SM, but the interactions between
Higgs bosons and quarks in the LH model are different with the
corresponding ones in the SM.

\par
The relative effect of the LH on the cross section
[$\delta=(\hat{\sigma}_{LH}-\hat{\sigma}_{SM})/\hat{\sigma}_{SM}$]
for subprocess $gg \to H^0H^0$ as functions of the c.m.s. energy
($\sqrt{\hat{s}}$) of incoming gluons for the cases of $x=0$
(corresponding to $M_{\Phi^0}=3.06~{\rm TeV}$) and $x=1/\sqrt{2}$
(corresponding to $M_{\Phi^0}=4.32~{\rm TeV}$) are depicted in
Fig.3(a), with the parameters taken as $f=3.5~{\rm TeV}$ and
$m_{H^0} =150~ {\rm GeV}$. In both figures, the solid line, dashed
line and dotted line correspond to $\lambda_1/\lambda_2=1/2$,
$\lambda_1/\lambda_2=1$, and $\lambda_1/\lambda_2=2$,
respectively. From Eq. (\ref{mt1})-(\ref{l1l2}) we can get that
those curves correspond to $m_T=6.28~{\rm TeV}$, $5~{\rm TeV}$,
and $6.28~{\rm TeV}$, separately. All the curves in Fig. 3(a) have
the common line structure which decreases rapidly in the vicinity
of the Higgs boson pair production threshold with an increment of
$\sqrt{\hat{s}}$, and then increases steadily after arriving at
its minimal value. The curves also obviously show that with an
increase of the value of $\lambda_1/\lambda_2$, the effect of the
LH model is getting stronger.

\par
In order to clarify the line shape in Fig. 3(a) more clearly, we
present the cross sections in the LH model ($\hat{\sigma}_{LH}$),
the SM ($\hat{\sigma}_{SM}$), and the difference between them
($\hat{\sigma}_{LH}-\hat{\sigma}_{SM}$) for the subprocess $gg \to
H^0H^0$ as functions of $\sqrt{\hat{s}}$ with $m_{H^0}=150~{\rm
GeV}$, $f=3.5~{\rm TeV}$, $\lambda_1/\lambda_2=1$, and  $x=0$ in
Fig. 3(b) and Table 1. The solid line, dashed line and dotted line
in Fig. 3(b) correspond to $\hat{\sigma}_{SM}$,
$\hat{\sigma}_{LH}$, and $\hat{\sigma}_{LH}-\hat{\sigma}_{SM}$,
respectively. We can see from Fig. 3(b) that in the
$\sqrt{\hat{s}}$ region from $310~{\rm GeV}$ to $500~{\rm GeV}$
the curve for $\hat{\sigma}_{SM}$ rises up steeply, while the
$\hat{\sigma}_{LH}-\hat{\sigma}_{SM}$ changes gently, which can be
also read from Table 1. Fig. 3(b) and Table.1 show that the
polelike behavior of the c.m.s. energy around $\sqrt{\hat{s}} \sim
400$-$500~{\rm GeV}$ in Fig. 3(a), comes from the fact that the
cross section($\hat{\sigma}_{SM}$) rises up steeply when
$\sqrt{\hat{s}}$ is just beyond the threshold energy and decreases
gently after $\hat{\sigma}_{SM}$ reaches its maximal value
$\sqrt{\hat{s}} \sim 500~{\rm GeV}$, while the variation of
$\hat{\sigma}_{LH}-\hat{\sigma}_{SM}$ is relatively slow in our
plotted energy range.
\begin{table}[htb]
\centering
\begin{tabular}{|l|c|c|c|c|}
\hline
$\sqrt{\hat{s}}$ & $\hat{\sigma}_{SM}$ & $\hat{\sigma}_{LH}$ & $\hat{\sigma}_{LH}-\hat{\sigma}_{SM}$ &
$(\hat{\sigma}_{LH}-\hat{\sigma}_{SM})/\hat{\sigma}_{SM}$\\
\hline
310  & 0.0052071 &  0.0060828 & 0.0008756 & 0.168162  \\
\hline
330  & 0.0268346 &  0.0295622 & 0.0027275 & 0.101643  \\
\hline
350  & 0.0884609 &  0.0947308 & 0.0062699 & 0.0708781 \\
\hline
370  & 0.2052696 &  0.2169084 & 0.0116388 & 0.0567002 \\
\hline
400  & 0.3425389 &  0.3595995 & 0.0170606 & 0.0498064 \\
\hline
420  & 0.4051675 &  0.4245253 & 0.0193578 & 0.0477772 \\
\hline
440  & 0.4476802 &  0.4685783 & 0.0208981 & 0.0466808 \\
\hline
460  & 0.4738619 &  0.4957345 & 0.0218726 & 0.0461582 \\
\hline
490  & 0.4903020 &  0.5128817 & 0.0225796 & 0.0460525 \\
\hline
500  & 0.4912129 &  0.5138794 & 0.0226666 & 0.0461441 \\
\hline
520  & 0.4880278 &  0.5107043 & 0.0226765 & 0.0464655 \\
\hline
540  & 0.4797802 &  0.5022973 & 0.0225171 & 0.0469321 \\
\hline
590  & 0.4462748 &  0.4679308 & 0.0216559 & 0.048526  \\
\hline
700  & 0.3565471 &  0.3754663 & 0.0189191 & 0.0530621 \\
\hline
900  & 0.2334130 &  0.2478193 & 0.0144063 & 0.06172   \\
\hline
1000 & 0.1935886 &  0.2063001 & 0.0127115 & 0.0656626 \\
\hline
2000 & 0.0598705 &  0.0659044 & 0.0060339 & 0.100783  \\
\hline
\end{tabular}
\caption{\small The cross sections of $gg \to H^0H^0$ in the LH
model and SM, the difference between them, and the relative
correction with $m_{H^0}=150~{\rm GeV}$, $f=3.5~{\rm TeV}$,
$\lambda_1/\lambda_2=1$, and  $x=1/\sqrt{2}$}.
\end{table}

\par
The dependence of the ratio $\hat{\sigma}_{LH}/\hat{\sigma}_{SM}$
of the subprocess $gg \to H^0H^0$ on the parameter $x$ with
$m_{H^0}=150~{\rm GeV}$ and $\sqrt{\hat{s}}=800~{\rm GeV}$, is
depicted in Fig. 4(a). The solid line, dashed line and dotted line
correspond to $\lambda_1/\lambda_2=1/2$, $\lambda_1/\lambda_2=1$,
and $\lambda_1/\lambda_2=2$, respectively. Fig. 4(a) shows that
the effect of the LH model always enhances the cross section of
$gg \to H^0H^0$ in our chosen parameter space. The three curves
also demonstrate that the effect of the LH model is not sensitive
to $x$ in the range of $x \lesssim 0.8$, but their values increase
rapidly and can be larger than 1.1 when $x > 0.85$ for all three
curves. To explain the result that the correction blows up as we
take the $x \to 1$ limit shown in Fig. 4(a), we decompose the
cross sections of subprocess $gg \to H^0H^0$ in the LH model into
three parts:
\begin{eqnarray}
\hat{\sigma}_{LH}=\hat{\sigma}_{box}+\hat{\sigma}_{tri}+\hat{\sigma}_{int},
\end{eqnarray}
where we denote $\hat{\sigma}_{box}$, $\hat{\sigma}_{tri}$, and
$\hat{\sigma}_{int}$ as the contribution parts from the box
diagrams[including quartic diagrams, shown in Fig. 1(2), 1(3),
1(5), 1(7)], the triangle diagrams[shown in Fig. 1(1), 1(4)] and
the interference between the box and triangle diagrams
respectively. In Fig. 4(b), we show the contributions of these
three parts as functions of $x$ with the conditions of
$\sqrt{\hat{s}}=800~{\rm GeV}$, $m_{H^0}=150~{\rm GeV}$,
$f=3.5~{\rm TeV}$ and $\lambda_1/\lambda_2=1$. The solid line,
dotted line, dash dotted line, and dashed line correspond to box
diagrams, triangle diagrams, interference contributions, and total
cross section, respectively. Our calculation demonstrates that the
main contribution to the cross section of subprocess $gg \to
H^0H^0$ comes from the box diagrams, and the interference
contribution is mainly from the Feynman diagrams involving the top
quark. Let us review the couplings of $H^0$-$H^0$-$H^0$ and
$H^0$-$\bar{t}$-$t$ in the LH model. They can be expressed as
\begin{eqnarray}
g_{H^0H^0H^0}&:&-i\left(\frac{3 m_{H^0}^2}{v} - \frac{33 m_{H^0}^2
v}{4 f^2}\frac{x^2}{1-x^2}\right), \\
g_{H^0\bar{t}t}~~~~~&:&-i\frac{m_t}{v} \left[1+\frac{x v^2}{2
f^2}-\frac{x^2 v^2}{4 f^2}-\frac{2}{3}\frac{v^2}{f^2}+
\frac{v^2}{f^2}\frac{\lambda_1^2}{\lambda_1^2+\lambda_2^2}
\left(1+
\frac{\lambda_1^2}{{\lambda_1}^2+{\lambda_2}^2}\right)\right].
\end{eqnarray}
We can see that the contribution of the triangle Feynman diagrams
is \textit{s}-channel suppressed and is relative small, due to the
heavy $\Phi^0$ and $\sqrt{\hat{s}}=800~{\rm GeV} >> m_{H^0}$. The
figure shows that the contribution from the interference between
the box diagrams and triangle diagrams is negative and blows up
quickly when we take the $x\to 1$ limit. We also find that the
dependence of the product of  $g_{H^0H^0H^0}$ and
$g_{H^0\bar{t}t}$ on the parameter $x$ behaves with the same rapid
increment when $x$ is close to 1. Therefore, we can conclude that
the quick enhancement behavior of the total cross section in the
vicinity where $x \to 1$ arises mainly from contributions of the
interference between the triangle diagrams involving the
$g_{H^0H^0H^0}$ vertex and the box diagrams involving the
$g_{h^0\bar{t}t}$ coupling. As we know, perturbativity alone
should put some bounds on the range that x should be allowed to
vary. But with these limitations some couplings in the original
Higgs potential must behave badly. Since we started with a
relatively will-behaved Higgs potential, it is clear that $x\to1$
cannot be a very well-defined limit and should not be considered
as a physical limit. In other words,the original Higgs potential
is an effective potential result of integrating out the heavy
states. Although it cannot give definitive limitations, the size
of all Higgs potential parameters should be roughly order 1
theoretically. We depict the relations between x (or $g_{HHH}$)
and the absolute values of original Higgs potential parameters
$\lambda_{\phi^2}$, $\lambda_{h\phi h}$, $\lambda_{h^4}$, and
$\lambda_{h\phi \phi h}$ [see Eq. (\ref{win1})] in Fig. 4(c) in
the conditions of $f=3.5~{\rm TeV}$ and $m_{H^0}=150~{\rm GeV}$.
In the figure, the solid line, dashed line, dotted line, and
dash-dotted line correspond to $|\lambda_{\phi^2}|$,
$|\lambda_{h\phi h}|$, $|\lambda_{h^4}|$, and $|\lambda_{h\phi
\phi h}|$, respectively. The absolute value of the coupling
$g_{H^0H^0H^0}$ corresponding to $x$ value is scaled on the upper
axis. From the figure, we can see that the values of all four
Higgs potential parameters, which are concerned in our
calculation, are far beyond order 1 when $x\gtrsim0.95$ and
$|\lambda_{h\phi \phi h}|$ is less than $10^{-1}$ when
$x\lesssim0.25$. Therefore, the above $x$ value range should be
excluded in our consideration, since these absolute values of the
Higgs potential parameters cannot satisfy the request of the
well-behaving effective Higgs potential. Then we can approximately
put bounds o the varying range of the parameter $x$ as
$0.25<x<0.95$ on the condition of $f=3.5~{\rm TeV}$ and
$m_{H^0}=150~{\rm GeV}$.

\par
The ratio of total cross sections $\sigma_{LH}/\sigma_{SM}$, as
functions of $x$ at the LHC ($\sqrt{s}=14~{\rm TeV}$) with
$f=3.5~{\rm TeV}$ and $\lambda_1/\lambda_2=2$ is depicted in Fig.
5. The total cross section involves the contributions of both
subprocesses $gg \to H^0H^0$ and $b \bar{b} \to H^0H^0$. In the
figure, the solid line, dashed line, and dotted line correspond to
$m_{H^0}=115~{\rm GeV}$, $150~{\rm GeV}$, and $180~{\rm GeV}$,
respectively. We can see again that the deviation of the total
cross section in the LH model from the corresponding SM value is
not sensitive to $x$ when $x\lesssim 0.8$, but increases quickly
when $x > 0.9$, and the relative deviation, defined as
$\delta=(\sigma_{LH}-\sigma_{SM})/\sigma_{SM}$, can be beyond
$12\%$ when $x$ reach $0.9$ at the LHC. We can see from this
figure that the correction effect of the LH model is obviously
related to the Higgs boson mass. When the Higgs boson mass varies
in the range of $115-180~{\rm GeV}$, the curves show that the
heavier the Higgs boson mass is, the stronger the correction
effect becomes.

\par
In Fig. 6 we plot the ratio between the total cross sections in
the LH model and the SM, as the functions of the mass of the
neutral Higgs boson $m_{H^0}$ with the conditions of $f=3.5~{\rm
TeV}$ and $\lambda_1/\lambda_2=2$ at the LHC. In this figure, the
solid line, dashed line, and dotted line correspond to the cases
of $x=0.25$, $1/\sqrt{2}$ and $0.9$ respectively. The relative
correction of the LH model to the SM cross section
[$\delta=(\sigma_{LH}-\sigma_{SM})/\sigma_{SM}$], can reach
$14.8\%$ when $x = 0.9$ at the LHC.

\par
The total cross sections $\sigma_{SM}$ and $\sigma_{LH}$ of
process $pp \to H^0H^0$ as functions of the mass of the neutral
Higgs boson $m_{H^0}$ is depicted in Fig. 7 at the LHC, with
$f=3.5~{\rm TeV}$, $x=0.9$, and $\lambda_1/\lambda_2=2$ in the LH
model. The solid line and dashed line correspond to the cases of
the SM and LH model, respectively. They show that the absolute
correction induced by the LH model decreases with an increase of
the Higgs boson mass.

\par
In Fig. 8, we plot the total cross sections $\sigma_{LH}$ of
process $pp \to H^0H^0$ as functions of the new heavy vector-like
quark mass $m_T$ with $m_{H^0}=150~{\rm GeV}$ and $x=1/\sqrt{2}$
in the LH model at the LHC. In this figure the solid line is for
$\lambda_1/\lambda_2=1/2$, dashed line for
$\lambda_1/\lambda_2=1$, and dotted line for
$\lambda_1/\lambda_2=2$. According to Eq. (\ref{mt2}) for a fixed
value of $\lambda_1/\lambda_2$, the mass of heavy-vector-like
quark $T$ is only related to the symmetry breaking scale parameter
$f$. For example, when $\lambda_1/\lambda_2=2$ and $m_T$ varies
from $1.5~{\rm TeV}$ to $4.0~{\rm TeV}$, we have that the
parameter $f$ changes from $0.85~{\rm TeV}$ to $2.26~{\rm TeV}$.
We can see from this figure that in the chosen parameter space the
cross sections are enhanced when either $m_T$ is relatively small
or $\lambda_1/\lambda_2$ is relatively larger.

\par
In conclusion, we investigated the effect of the LH model on the
pair production process of neutral Higgs bosons via both gluon and
bottom fusions at the LHC. The numerical analysis shows that with
the possible parameters, the relative cross section correction to
the SM prediction may reach a value of $24\%$ at the LHC when
$x=0.95$. We conclude that when the parameter $x$ has a value
above 0.8, the relative correction contributed by the LH model
reached a value beyond $8\%$ and could be observable at the LHC.

\vskip 10mm \noindent{\large\bf Acknowledgement:} This work was
supported in part by the National Natural Science Foundation of
China and a grant from the University of Science and Technology of
China.

\vskip 5mm \noindent{\large {\bf Appendix}}

\par
The interactive Lagrangian of the scalar field $\Sigma$ and the
up-type quarks of the first two generations take the same form as
in Eq. (\ref{L1}), except that there is no need for the extra
$\tilde{t}$. The interactive Lagrangian of the $\Sigma$ and the
down-type quarks can be expressed as
\begin{eqnarray}
{\cal L}_Y =\frac{1}{2}\lambda_d f
\epsilon_{ijk}\epsilon_{xy}\chi_i\Sigma_{jx}^*\Sigma_{ky}^*d^c+H.C.,
\end{eqnarray}
where the isospin index $i=1,2$. All the Yukawa couplings between
the Higgs boson and quarks (except the Yukawa coupling which
involves the top quark) can be obtained from the corresponding
Lagrangian directly. We present the expressions of the
$H^0$-$H^0$-$\bar{u}$-$u$ and $H^0$-$H^0$-$\bar{d}$-$d$ couplings
in the equations
\begin{eqnarray}
\label{hhqq} g_{H^0H^0\bar{d}d}~: ~-i\frac{4 m_d v'}{v^2(f +
v')}~~~,~~~ g_{H^0H^0\bar{u}u}~: ~-i\frac{4 m_u v'}{v^2(f + v')}~.
\end{eqnarray}
The expressions of other couplings concerned with in this work can
be found in Ref. \cite{LH8}.

\par
In the LH model, the trilinear interaction of Higgs bosons
$g_{H^0H^0H^0}$ gets a correction to the SM at the order of
$v^2/f^2$, and an additional $H^0H^0\Phi^0$ coupling is generated.
They are given by the Lagrangians
\begin{eqnarray}
-{\cal L}_{HHH}
&=& H^0 H^0 H^0 \left(-\frac{2 f v'\lambda_{h\phi h}}{v} +
  v\lambda_{h^4} - \frac{12~ v'^2 \lambda_{h^4}}{v} + \frac{6 ~v'^2 \
\lambda_{h\phi\phi h}}{v}\right),\nonumber \\
-{\cal L}_{HH\phi}
&=&H^0 H^0 \phi^0 \left(-\frac{f \lambda_{h\phi
h}}{\sqrt{2}} + \frac{14 \sqrt{2}~f v'^2 \ \lambda_{h\phi h}}{v^2}
-6\sqrt{2}~v'\lambda_{h^4} + \frac{5 v' \lambda_{h\phi\phi
h}}{\sqrt{2}}\right).
\end{eqnarray}
From the above Lagrangians we obtain the Feynman rule for
$H^0H^0H^0$ and $H^0H^0\Phi^0$ couplings as \cite{LH8}
\begin{eqnarray}
g_{H^0H^0H^0}&:&-i\left(\frac{3 m_{H^0}^2}{v} - \frac{66 M_{\Phi^0}^2 v'^2}{f^2 v}\right), \nonumber \\
g_{H^0H^0\phi^0}&:& -i\left( \frac{56\sqrt{2}M_{\Phi^0}^2
v'^3}{v^4}- \frac{2 \sqrt{2} M_{\Phi^0}^2 v'}{v^2} -\frac{29
\sqrt{2}~M_{\Phi^0}^2 v'}{3 f^2}\right).
\end{eqnarray}

\vskip 10mm
\begin{flushleft} {\bf Figure Captions} \end{flushleft}
\par
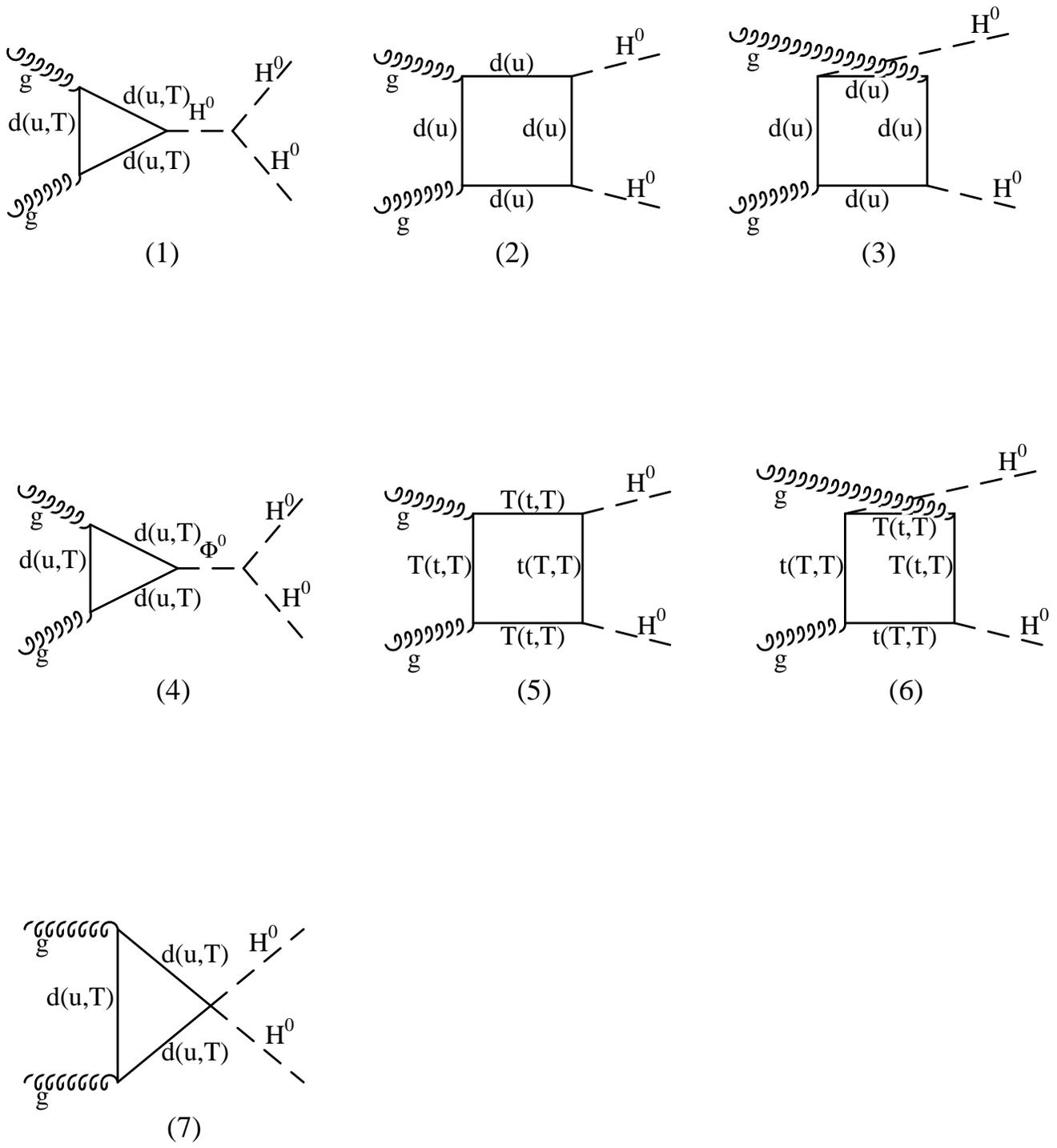
{\bf Fig. 1} The one-loop Feynman diagrams of the subprocess $gg
\to H^{0}H^{0}$ in the LH model: (1), (4) \textit{s}-channel
diagrams. (2), (3), (5), (6) box diagrams. (7) quartic diagrams.
The notations $u$ and $d$ represent up-type and down-type quarks,
respectively. The diagrams created by exchanging two external
gluon lines and final Higgs boson lines are not shown.

\par
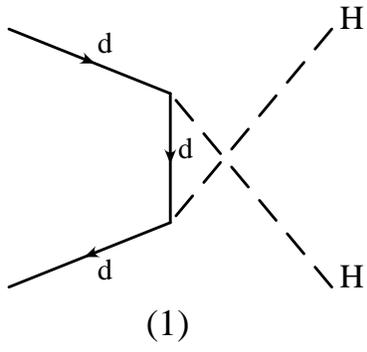
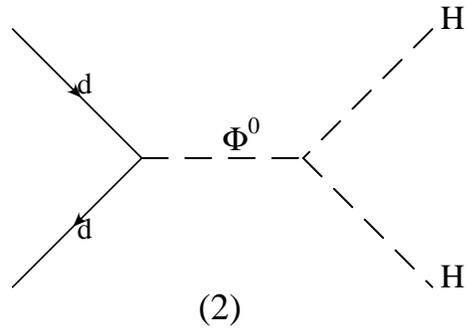
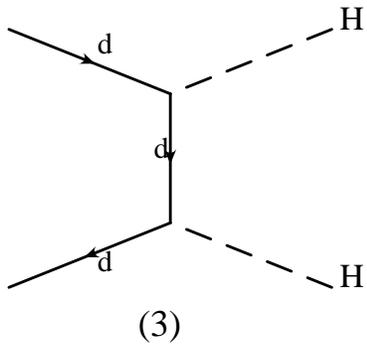
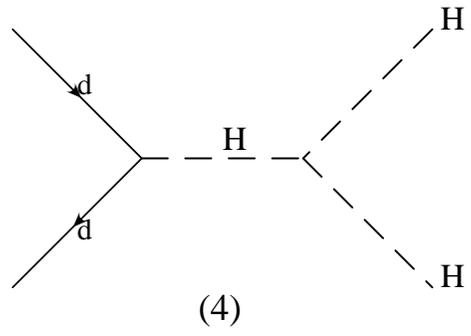
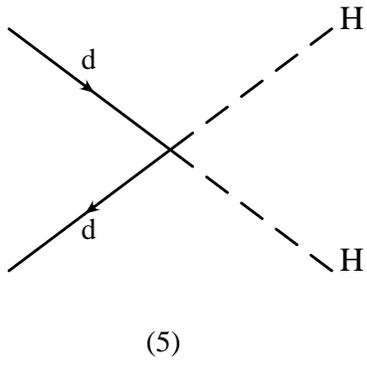
{\bf Fig. 2} The lowest-order Feynman diagrams of the subprocess
$b\bar{b} \to H^{0}H^{0}$ in the LH model.

\par
{\bf Fig.3(a)} The relative effect of the LH on the cross section
[$\delta=(\hat{\sigma}_{LH}-\hat{\sigma}_{SM})/\hat{\sigma}_{SM}$]
for subprocess $gg\to H^0H^0$ as functions of $\sqrt{\hat{s}}$ on
the conditions of $x=1/\sqrt{2}$,$f=3.5~{\rm TeV}$, $m_{H^0} =150
~{\rm GeV}$. The solid line, dashed line, and dotted line
correspond to $\lambda_1/\lambda_2=1/2$, $\lambda_1/\lambda_2=1$,
and $\lambda_1/\lambda_2=2$, respectively. {\bf(b)} The cross
sections of $gg\to H^0H^0$ subprocess in the SM, the LH model, and
their difference $\hat{\sigma}_{LH}-\hat{\sigma}_{SM}$ as
functions of $\sqrt{\hat{s}}$ with $x=1/\sqrt{2}$, $f=3.5~{\rm
TeV}$, $m_{H^0} =150 ~{\rm GeV}$, and $\lambda_1/\lambda_2=1$.

\par
{\bf Fig. 4(a)} The dependence of the ratio
$\hat{\sigma}_{LH}/\hat{\sigma}_{SM}$ for the subprocess $gg \to
H^0H^0$ on the parameter $x$ with $m_{H^0}=150~{\rm GeV}$ and
$\sqrt{\hat{s}}=800~{\rm GeV}$. The solid line, dashed line and
dotted line correspond to $\lambda_1/\lambda_2=1/2$,
$\lambda_1/\lambda_2=1$, and $\lambda_1/\lambda_2=2$,
respectively. {\bf (b)}The total cross section and contributions
of the box diagrams (including quartic diagrams), the triangle
diagrams, interference between the box and triangle diagrams
($\hat{\sigma}_{LH}=\hat{\sigma}_{box}+\hat{\sigma}_{tri}+
\hat{\sigma}_{int}$) for the subprocess $gg \to H^0H^0$, as
functions of $x$ with the conditions of $\sqrt{\hat{s}}=800~{\rm
GeV}$, $m_{H^0}=150~{\rm GeV}$, $f=3.5~{\rm TeV}$, and
$\lambda_1/\lambda_2=1$. The solid line, dotted line, dash dotted
line, and dashed line correspond to box diagrams, triangle
diagrams, interference contributions, and total cross section
respectively. {\bf (c)}The absolute value of the original Higgs
potential parameters $\lambda_{\phi^2}$, $\lambda_{h\phi h}$,
$\lambda_{h^4}$ and $\lambda_{h\phi \phi h}$ as functions of $x$
with the conditions of $m_{H^0}=150~{\rm GeV}$ and $f=3.5~{\rm
TeV}$. The absolute value of coupling $g_{H^0H^0H^0}$
corresponding to the $x$ value is scaled on the upper axis.

\par
{\bf Fig. 5} The ratio of the total cross sections in the LH model
and SM $\sigma_{LH}/\sigma_{SM}$ for the process $pp \to H^0H^0$
at the LHC, as functions of $x~(=4fv'/v^2)$, with $f=3.5~{\rm
TeV}$ and $\lambda_1/\lambda_2=2$.

\par
{\bf Fig. 6} The ratio of the total cross sections
$\sigma_{LH}/\sigma_{SM}$ for the process $pp \to H^0H^0$ at LHC,
as functions of the Higgs boson mass $m_{H^0}$ with the conditions
of $f=3.5~{\rm TeV}$ and $\lambda_1/\lambda_2=2$ in the LH model.

\par
{\bf Fig.7} The total cross sections $\sigma_{SM}$ and
$\sigma_{LH}$ for the process $pp \to H^0H^0$ at the LHC, as
functions of the Higgs boson mass $m_{H^0}$ with $f=3.5~{\rm
TeV}$, $x=0.9$ and $\lambda_1/\lambda_2=2$.

\par
{\bf Fig.8} The total cross section $\sigma_{LH}$ as functions of
the mass of the new heavy-vector-like quark $m_T$ with
$m_{H^0}=150~{\rm GeV}$ and $x=1/\sqrt{2}$ in the LH model at the
LHC.
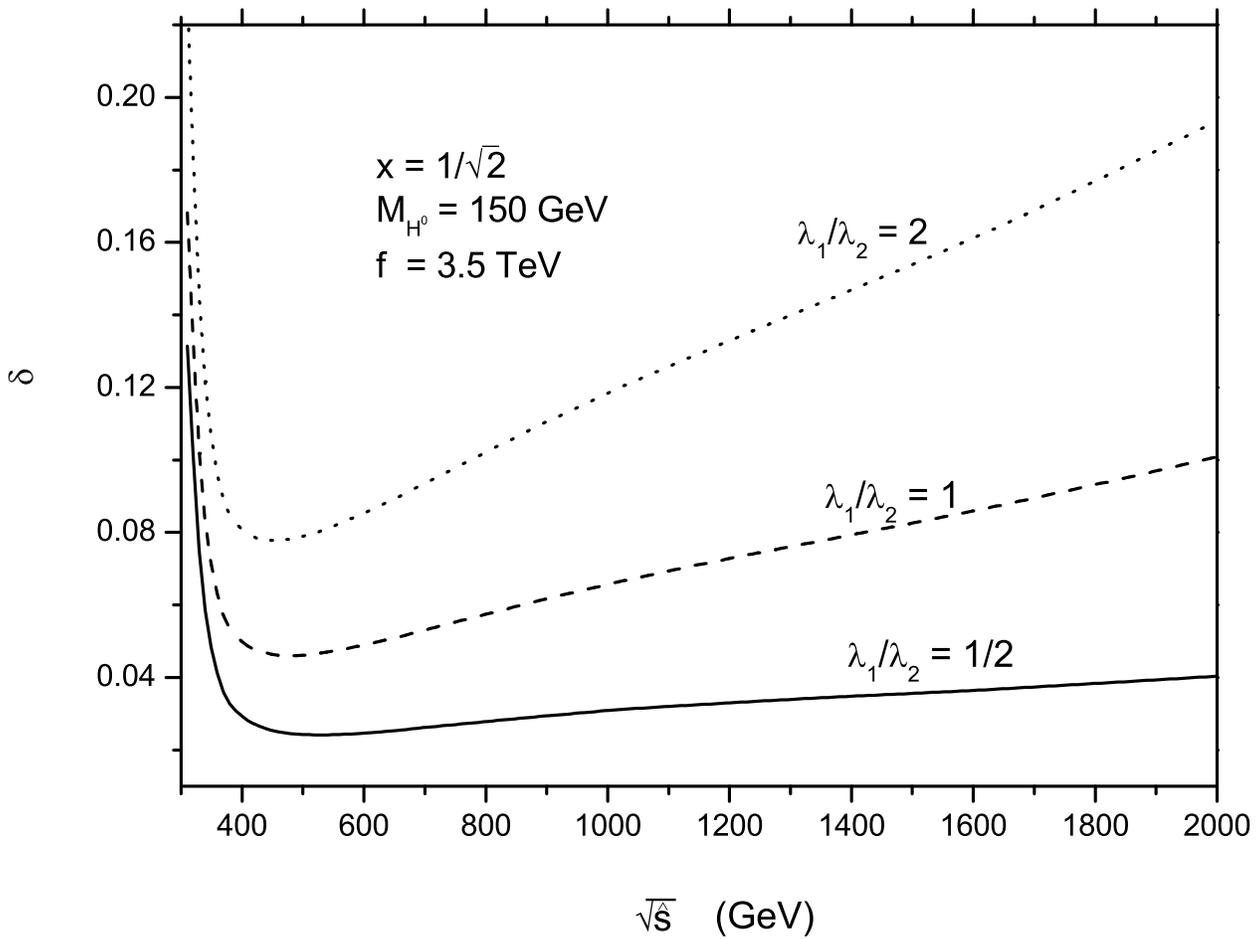
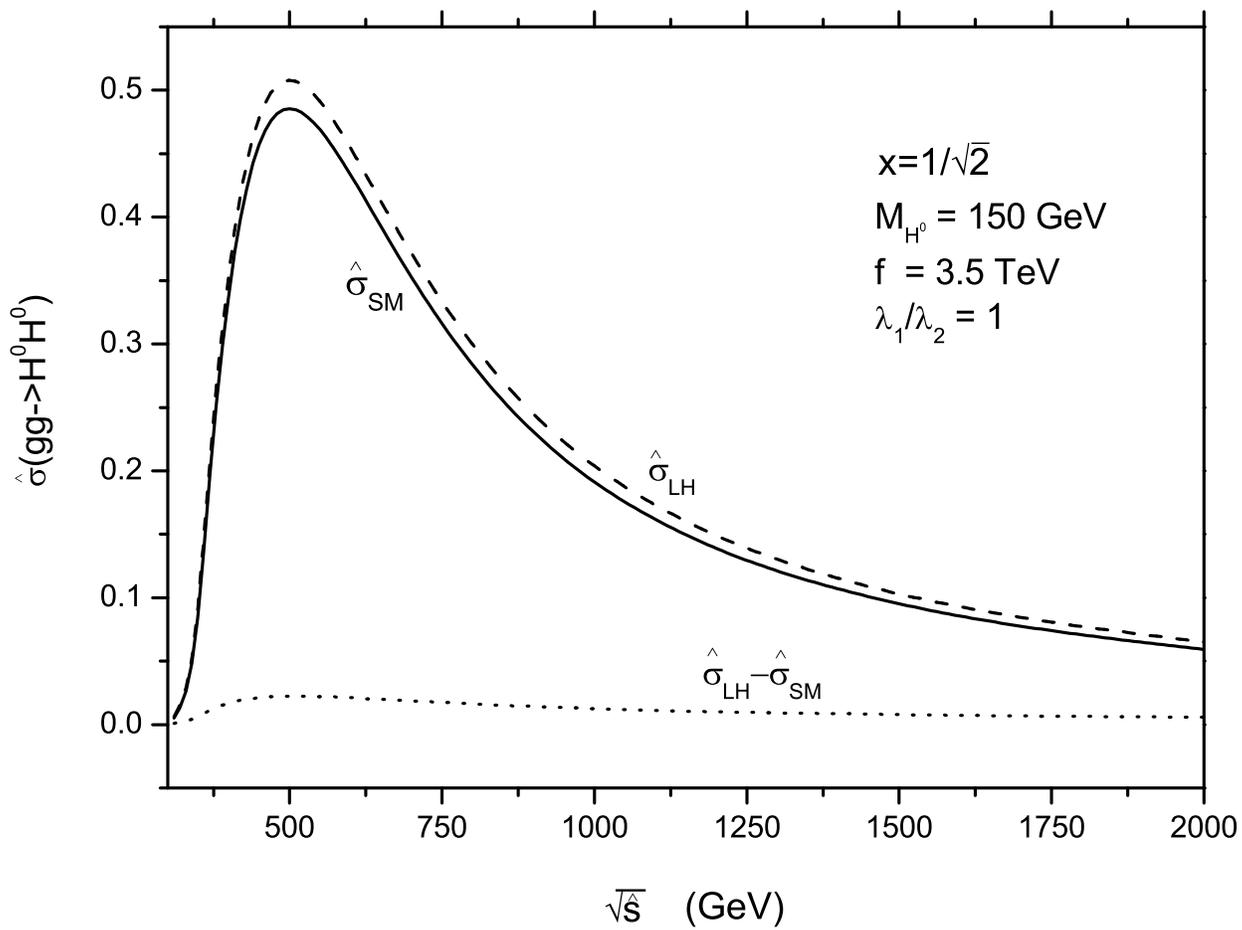
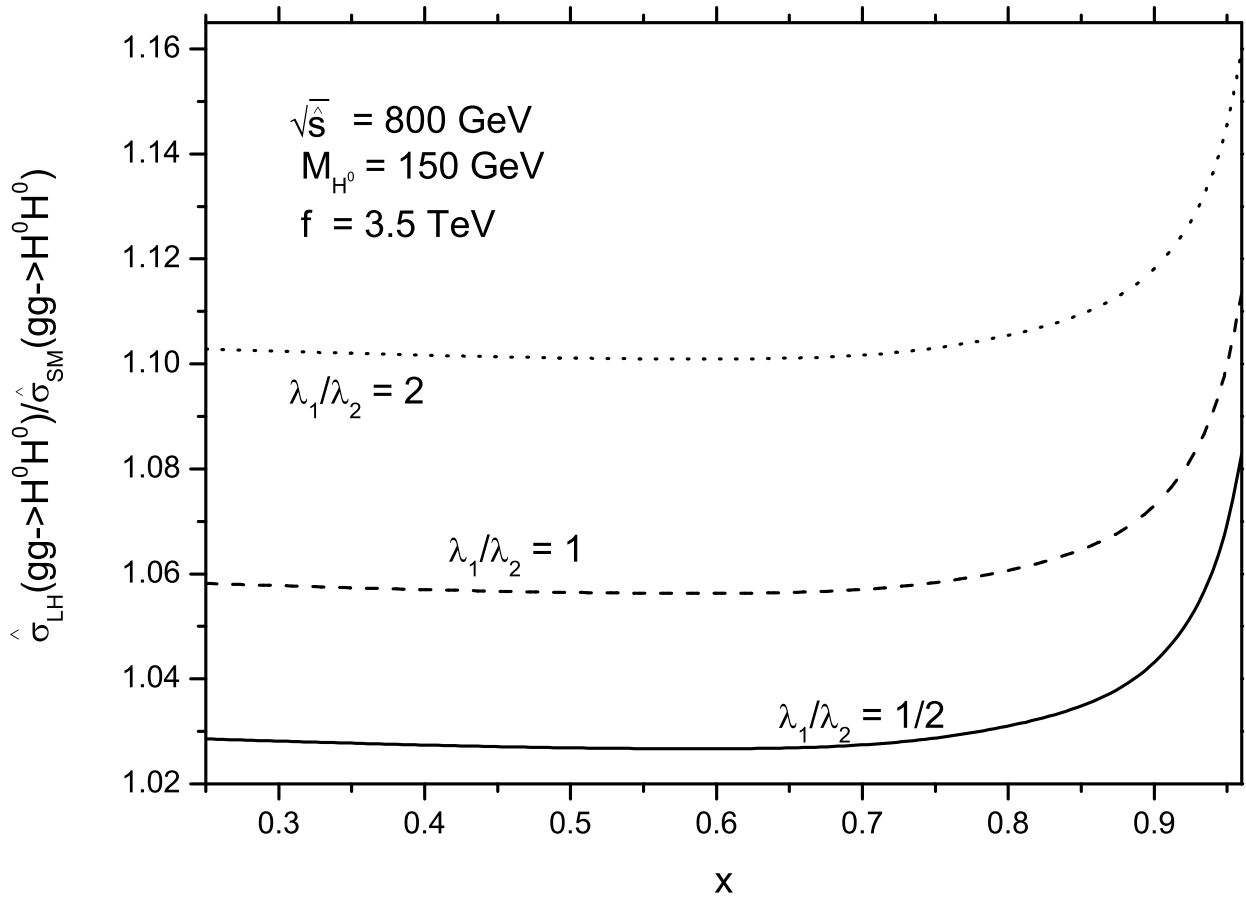
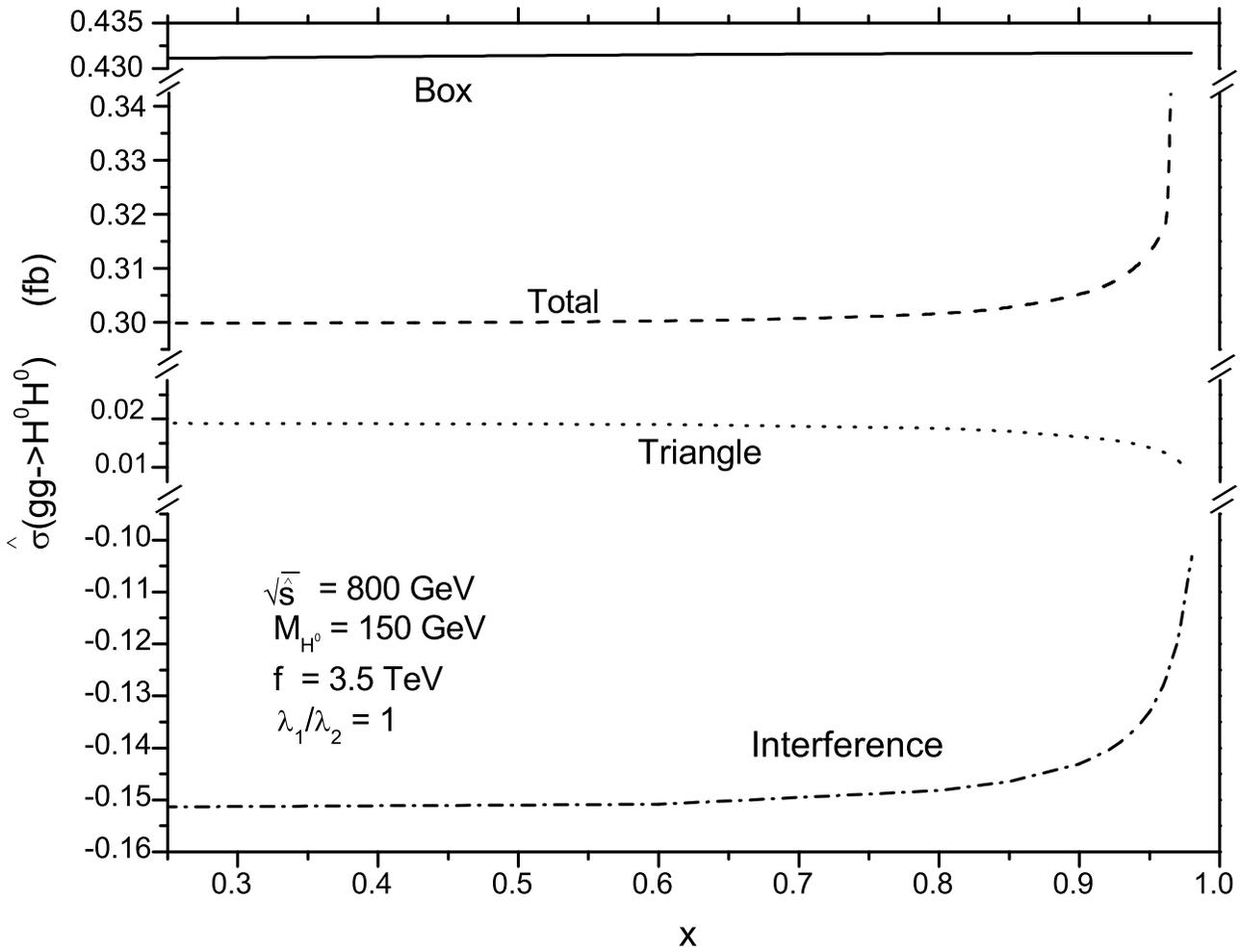
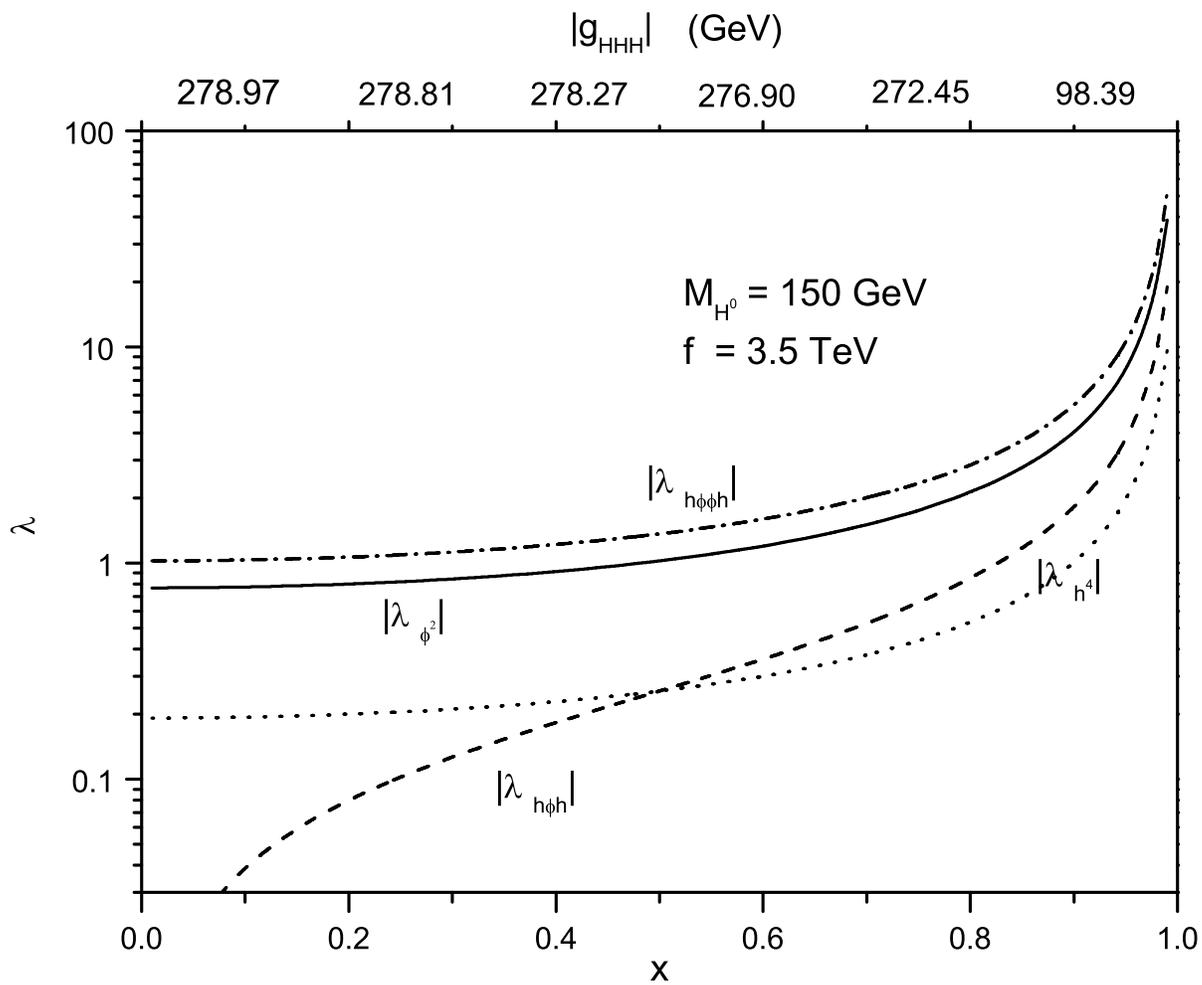
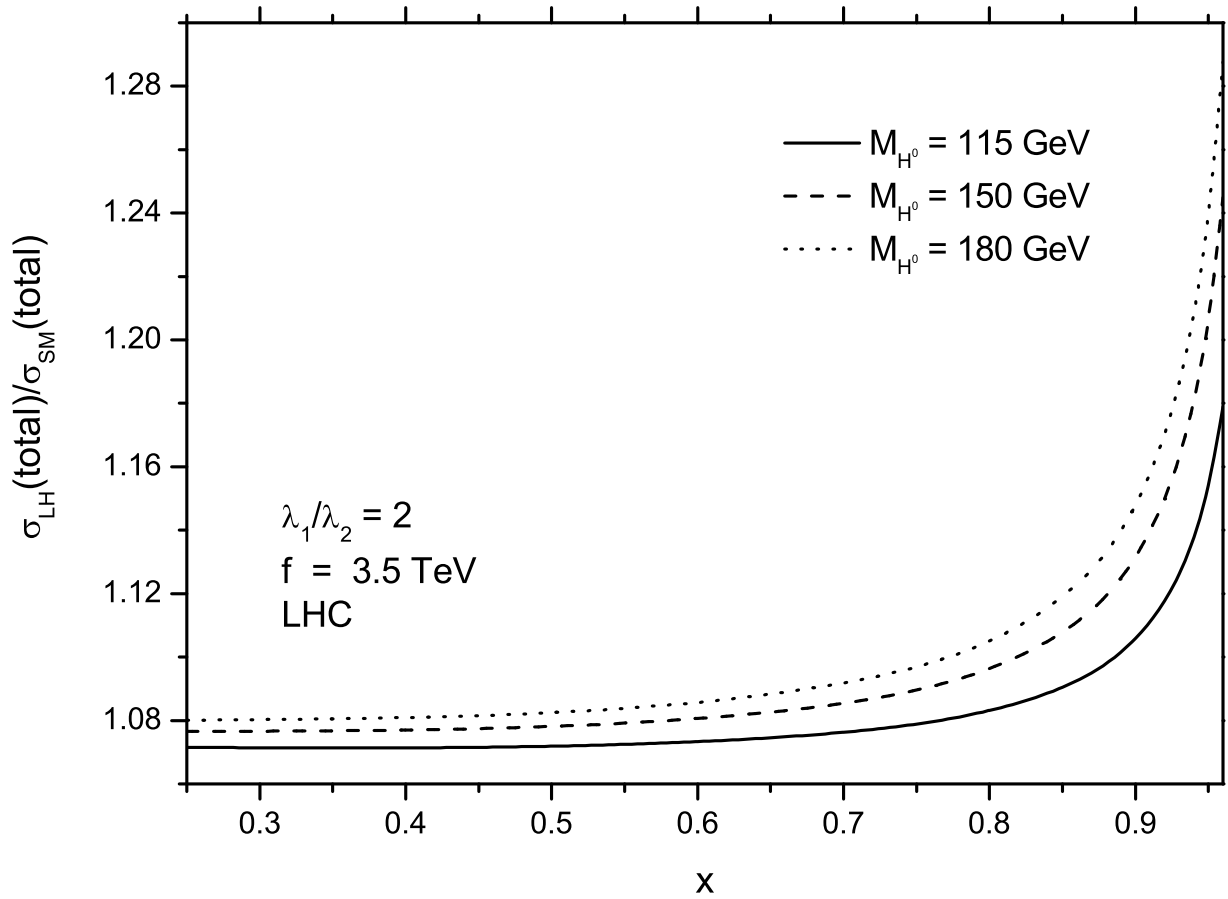
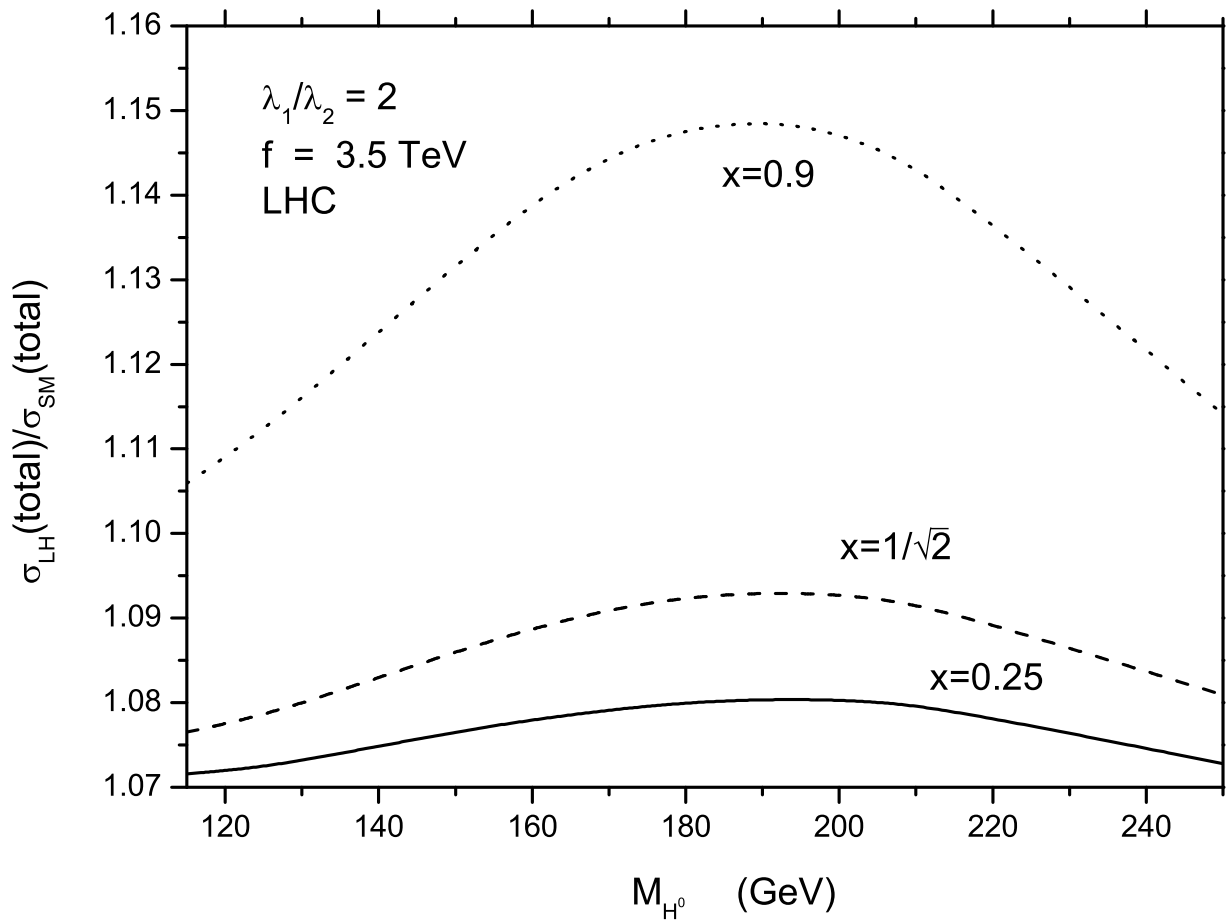
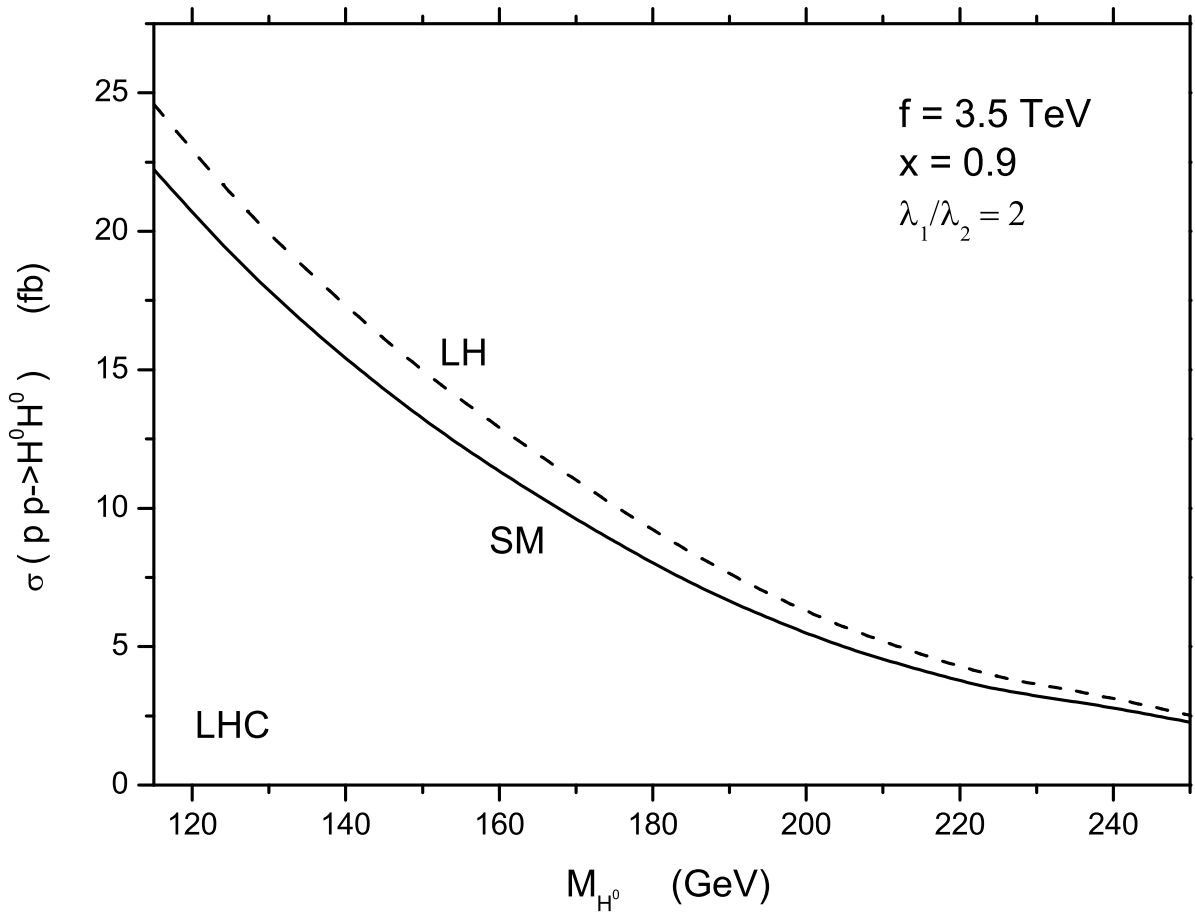
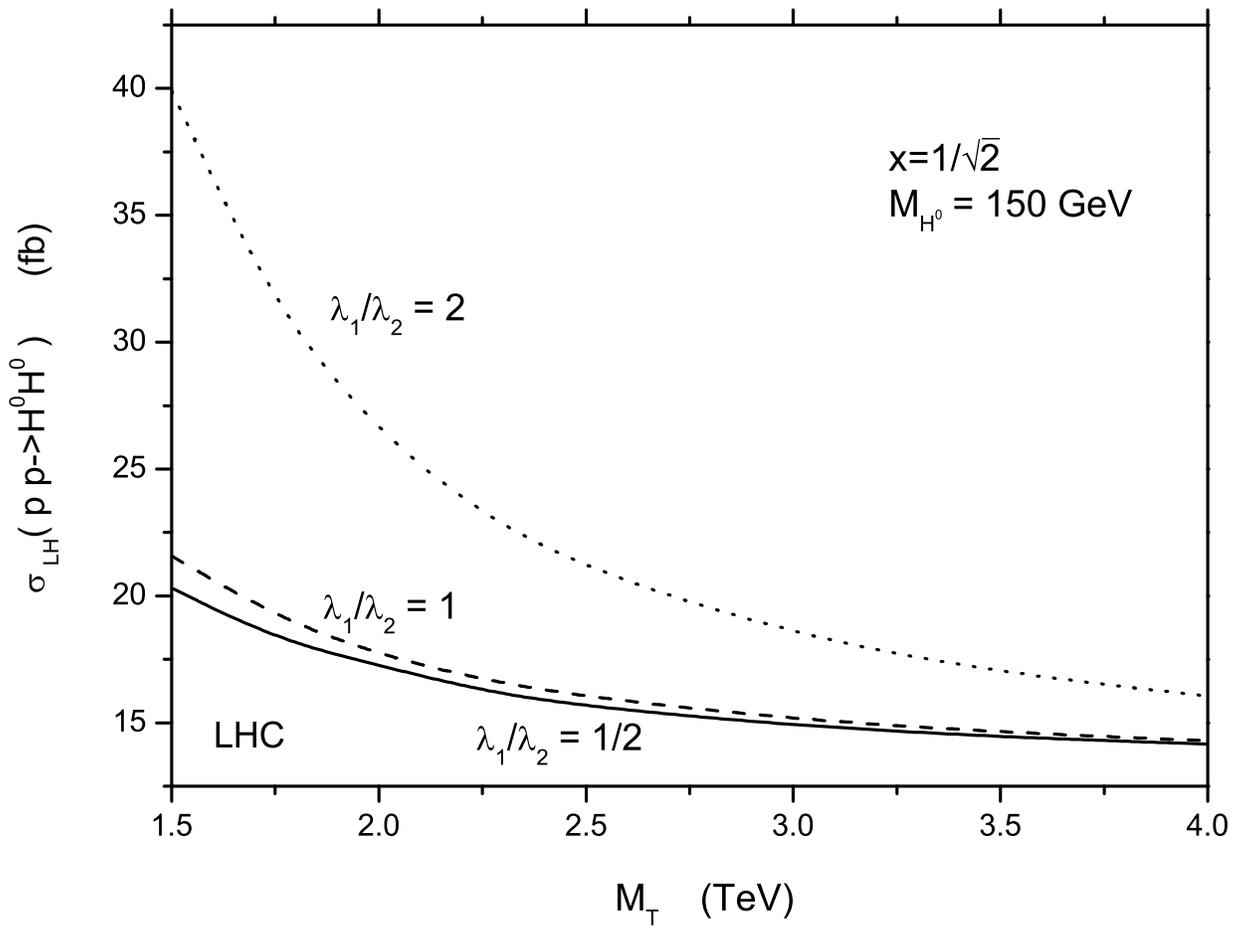
\end{document}